\documentclass[a4paper,twocolumn,british,prl,superscriptaddress,longbibliography,notitlepage]{revtex4-2}
\usepackage[T1]{fontenc}
\usepackage{amsmath}
\usepackage{amssymb}
\usepackage{graphicx}

\makeatletter

\pdfpageheight\paperheight
\pdfpagewidth\paperwidth

\usepackage{xcolor}
\definecolor{darkblue}{rgb}{0.1,0.2,0.6} 
\definecolor{lightblue}{rgb}{0.1,0.1,1.0}
\definecolor{darkred}{rgb}{0.8,0.1,0.2}
\usepackage[colorlinks,citecolor=lightblue,linkcolor=darkblue,urlcolor=lightblue] {hyperref}

\makeatother

\usepackage{babel}
\begin{document}
\global\long\def\bra#1{\left\langle #1\right|}%
\global\long\def\ket#1{\left|#1\right\rangle }%

\title{Discrete time-crystals in unbounded potentials}
\author{Yevgeny Bar~Lev}
\affiliation{Department of Physics, Ben-Gurion University of the Negev, Beer-Sheva
  84105, Israel}
\author{Achilleas Lazarides}
\affiliation{Loughborough University, Loughborough, Leicestershire LE11 3TU, UK}
\begin{abstract}
  Discrete time crystalline (DTC) phases have attracted significant
  theoretical and experimental attention in the last few years. Such
  systems require a seemingly impossible combination of non-adiabatic
  driving and a finite-entropy long-time state, which, surprisingly, is
  possible in non-ergodic systems. Previous works have often relied on
  disorder for the required non-ergodicity; here, we describe the
  construction of a discrete time crystal (DTC) phase in
  non-disordered, non-integrable Ising-type systems. After discussing
  the conditions for interacting and periodically driven systems to
  display such phases in general, we propose a concrete model and then
  provide approximate analytical arguments and direct numerical
  evidence that it satisfies the conditions and displays a DTC phase
  robust to local periodic perturbations.
\end{abstract}
\maketitle
\emph{Introduction}.---The study of dynamical phases of matter has been one of the most fruitful directions in many-body physics over the last few years. Time crystals, in which both spatial and temporal symmetries are spontaneously broken, constitute one of the most recognisable examples. Conceptually, these are attractive for two reasons: Firstly, they are genuine out of equilibrium phases of matter, impossible at equilibrium~\citep{watanabe2015:AbsenceQuantumTime}. Secondly, they surf the very edge of the second law of thermodynamics: They simultaneously require a system that is non-adiabatically perturbed, which would usually result in entropic increase (heating)~\citep{Lazarides:2015jd,Ponte:2015hm,DAlessio:2014fg}, but at the same time the system must not heat up to a featureless infinite temperature state, since then no spatiotemporal structures will emerge. These apparently contradictory requirements may be simultaneously satisfied in a way that we will explain in this work, and that was first used in a different, disordered system earlier \citep{Khemani2016}.

So far, most DTCs have been found in systems that are temporally perturbed, usually periodically (``Floquet systems''). Such systems generically heat up towards a uniform, featureless state unfavourable to nontrivial effects~\citep{Lazarides:2015jd,Ponte:2015hm,DAlessio:2014fg}. Obtaining nontrivial effects therefore requires consideration of finite-time properties (prethermal physics~\citep{Berges2004,Mori2017,Abanin2015a,Else2017}) or non-ergodic systems which do not heat up when driven.

The first approach taken to break ergodicity was using disorder~\citep{Abanin:2018tf,Nandkishore2014}.  It allowed for driving without heating, and is often called Floquet Many-Body Localization (Floquet-MBL)~\citep{Lazarides:2015jd,Ponte:2015dc,Rehn:2016ws}.  Not long afterwards Discrete Time Crystal (DTC) phases were proposed~\citep{Khemani2016,Else:2016ue,Yao2017} and observed~\citep{Zhang:2017ci,choi2017,Ho:2017ea,Mi2022a,Pal2018DTC} in Floquet systems.

It was later realised that a tilted potential (uniform force) results in localisation and therefore ergodicity breaking in certain interacting systems~\citep{schulz2019stark,van2019bloch,Ribeiro2020}. This effect, which is closely related to Bloch oscillations, has been known for many years for noninteracting systems~\citep{zener_theory_1934,Wannier1960}. For interacting systems the situation is somewhat more complicated than expected, since any \emph{finite} system with a \emph{purely} reflection-symmetric (such as linear) potential is \emph{not} localised~\citep{zisling2022:TransportStarkManybody,Kloss2022}. It has already been established that such systems remain localised under periodic driving~\citep{Bhakuni2020,deger2023stark}, and show certain features of DTC phase~\citep{Kshetrimayum2020b,Liu2023}.

In this Letter we show how to construct a DTC using a unbounded potentials to prevent ergodicity and the system's heat death in the thermodynamic limit. We provide numerical evidence for our claims for the linear (Stark) and third-order polynomial potentials as special cases.

We start by discussing the conditions for a system to be a time-liquid, a phase where in the thermodynamic limit and after infinitely long time the state has the same temporal symmetry as the Hamiltonian,  and conclude that  this is the default behaviour for generic systems. This includes most non-ergodic systems. We then identify two other possible phases: time-glasses, where the temporal symmetry is completely broken in the long-time steady state, and time-crystals, where the symmetry is reduced to a subgroup. We explain how to construct a time crystal starting from a time liquid, discussing in detail the spectral features required for stability. Finally, putting these to practice, we write down an Ising-type driven model, then show both analytically and numerically that it satisfies our criteria and forms a time crystal.


\begin{figure} \includegraphics[width=1\columnwidth]{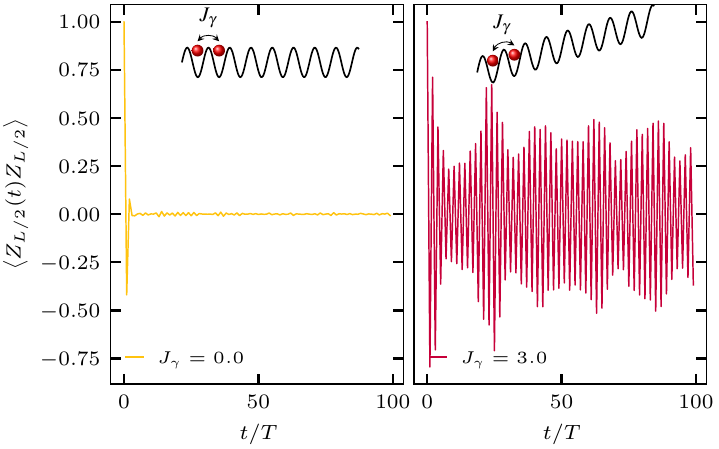}

  \caption{\label{fig:short-time-dynamics}Short-time dynamics of a local correlation function $\left\langle \hat{Z}_{L/2}\left(t\right)\hat{Z}_{L/2}\right\rangle $ for constant interactions (left), and tilted interactions (right) between the nearby spins. The parameters used for this simulation are $L=10$, $h=0.5$, $J_{0}=1$, $V=0.5$, $\theta/\pi=-0.1$ and $T=1$.} \end{figure}

\emph{Time-liquids}.---We define time-liquids (TL) as systems with a static or periodic Hamiltonians whose local observables approach a constant or a periodic function, correspondingly, at \emph{the   thermodynamic limit} and \emph{after infinitely long time }\citep{Reimann:2008hq,Linden:2009ii,Russomanno:2012bf,Lazarides:2014cl,Lazarides:2014ie,Ponte:2015hm} \footnote{It is easy to construct \emph{non-local} observables, with   arbitrary long-time dependence.}. That is, their long time dynamics has the same time-translation symmetry as the Hamiltonian \footnote{Similarly to a liquid which is invariant to spatial translations.}. The thermodynamic limit is essential since the observable in any \emph{finite} \emph{closed} system will fluctuate in time, while the infinite time limit is necessary because the initial condition itself breaks time-translation symmetry so that, for any \emph{finite} time, the temporal dependence of an observable cannot be expected to be time-translationally invariant.

For static time-liquids, the stationary value of an observable corresponds to its infinite-time average, $\overline{O}=\lim_{\tau\to\infty}\lim_{L\to\infty}\frac{1}{\tau}\int_{0}^{\tau}\mathrm{d}\bar{t}\left\langle \hat{O}\left(\bar{t}\right)\right\rangle $, where $L$ is the size of the system. For Floquet time-liquids we  focus on stroboscopic evolution. Stroboscopically, local observables in these systems will also approach a constant, $\overline{O}=\lim_{N\to\infty}\lim_{L\to\infty}\frac{1}{N}\sum_{n=1}^{N}\left\langle \hat{O}\left(nT\right)\right\rangle $, where $T$ is the period of the drive \footnote{The observable will generically change within the period.}. If the temporal fluctuations around the stationary value vanish in the thermodynamic limit, the system is a time-liquid.

We now proceed by examining the physical conditions for a system to be a time-liquid, focussing on the temporal fluctuations of a two-point correlation function of local observables,
\begin{equation}
  \Delta_{O_{i}O_{j}}^{2}\equiv\overline{\left|\left\langle \hat{O}_{i}\left(t\right)\hat{O}_{j}\right\rangle _{\rho}-\overline{\left\langle \hat{O}_{i}\left(t\right)\hat{O}_{j}\right\rangle _{\rho}}\right|^{2}},
  \label{eq:deltaO-Oflucts}
\end{equation}
where $\hat{O}_{i}$ is a local observable living in the vicinity of site $i$, $\left\langle \hat{O}\right\rangle _{\rho}\equiv\text{Tr }\hat{\rho}\hat{O}$ is the expectation with respect to the density matrix $\hat{\rho}$, and over-bar indicates an infinite-time average \footnote{We use the two-point correlation function and not $\left\langle \hat{O}_{i}\left(t\right)\right\rangle _{\rho}$ because the correlation function has a nontrivial time-dependence for all density matrices, $\hat{\rho}$. This includes density matrices which are invariant with respect to the dynamics, such as, thermal states for static systems.}. In what follows, we discuss under which conditions $\Delta_{O_{i}O_{j}}^{2}$ vanishes in the thermodynamic limit making the system a time liquid.

For a \emph{finite} system, temporal fluctuations, $\Delta_{O_{i}O_{j}}^{2}$, can be rigorously bounded: Assuming that the generator of the unitary evolution \footnote{For static systems the Hamiltonian, $\hat{H}$, while for periodically driven systems the effective Hamiltonian, $\hat{H}_{\text{eff}}=i\log\hat{U}_{T}$, where $\hat{U}_{T}$ is the one period unitary propagator.} has eigenvalues $E_{\alpha}$ with non-degenerate gaps~\citep{Reimann:2008hq,alhambra2020:TimeEvolutionCorrelationa,lezama2023temporal} so that $E_{\alpha}-E_{\beta}=E_{\alpha'}-E_{\beta'}$ if and only if $\alpha=\alpha'$ and $\beta=\beta'$ \footnote{While non-interacting systems do not satisfy this assumption, it typically holds for interacting systems. } the temporal fluctuations satisfy~\citep{lezama2023temporal}
\begin{equation}
  \Delta_{O_{i}O_{j}}^{2}\leq\left\Vert \hat{O}_{i}\right\Vert ^{2}\left\Vert \hat{O}_{j}\right\Vert ^{2}\sqrt{\operatorname{Tr}\bar{\rho}_{s}^{2}}.\label{eq:fluctuation-bound}
\end{equation}
Here $\left\Vert \hat{O}_{i}\right\Vert ^{2}$ is the operator norm of $\hat{O}_{i}$, which for local operators and bounded local Hilbert spaces is independent of the system size, and $\bar{\rho}_{s}\equiv\lim_{T\to\infty}\frac{1}{T}\int_{0}^{T}\mathrm{d}\bar{t}\,\hat{U}\left(t,0\right)\hat{\rho}\hat{U}\left(0,t\right)$ is the stationary state \footnote{For Floquet systems the infinite time average should be correspondingly adjusted}. We have $\bar{\rho}_{s}=\sum_{\alpha}\hat{P}_{\alpha}\hat{\rho}\hat{P}_{\alpha}$ with $\hat{P}_{\alpha}$ the projectors on the degenerate subspaces corresponding to the eigenvalue $E_{\alpha}$. $\bar{\rho}_{s}$ is typically a mixed state. Therefore if the purity of the stationary state vanishes in the thermodynamic limit, the system is a time-liquid.

For \emph{thermal} initial states, $\hat{\rho}=e^{-\beta\hat{H}}/Z=\bar{\rho}_{s}$. The purity of $\hat{\rho}$ can be bounded from above by $\exp\left(-S_{\beta}\right)$, with $S_{\beta}$ the (typically extensive) thermal entropy. Therefore, $\Delta_{O_{i}O_{j}}^{2}$ vanishes in the thermodynamic limit for all \emph{static} systems at thermal equilibrium, implying that \emph{static thermal systems are time liquids}. This constitutes a simple demonstration of the no-go theorem of Ref.~\citep{watanabe2015:AbsenceQuantumTime}. Even for \emph{nonthermal but macroscopic} systems, experimentally accessible initial states will typically involve a macroscopically large number of eigenstates~\cite{Reimann:2008hq} so that the system will be a time-liquid. This remains true even in the case of many non-ergodic systems. For example, temporal fluctuations of local observables in a many-body localized system (MBL) also vanish in the thermodynamic limit~\citep{serbyn2014:QuantumQuenchesManybody}. So, how can we avoid this fate?

\emph{Time glasses and time crystals}.---These phases are characterized by breaking of the time-translation symmetry of the unitary dynamics generator. Time-glasses break the symmetry completely while time-crystals reduce the symmetry to a symmetry subgroup. A number of definitions of these phases have been proposed~\citep{watanabe2015:AbsenceQuantumTime,khemani2017:DefiningTimeCrystals}. In this Letter we define a system to be a time crystal (glass) if a local observable exists, whose fluctuations around the stationary value starting from any physically realizable initial state are periodic (aperiodic). Moreover, we require that the phase will be stable to local, possibly, time-periodic perturbations. In what follows we only focus on the TC phase and propose a simple protocol for its construction.

\emph{Time-crystal construction}.---We consider a periodically driven system of \emph{interacting} spins with a unitary one-period propagator $\hat{U}_{0}\left(T\right)$, where $T$ is the driving period. We assume that $\hat{U}_{0}\left(T\right)$ commutes with the spin-flip operator, $\hat{P}=\prod_{j}\hat{X}_{j}$, where $\hat{X}_{j}$ are the $\hat{\sigma}_{j}^{x}$ Pauli matrices. We consider a one period propagator of the form, $\hat{U}\left(T\right)=\hat{P}\hat{U}_{0}\left(T\right)$. Using $\left[\hat{U}_{0}\left(T\right),\hat{P}\right]=0$, the unitary evolution of the correlation function is
\begin{align}
  C_{ij}\left(nT\right) & =\text{Tr }\left(\hat{\rho}\,\left(\hat{U}^{\dagger}\left(T\right)\right)^{n}\hat{Z}_{i}\left(\hat{U}\left(T\right)\right)^{n}\hat{Z}_{j}\right)\nonumber                       \\
                        & =\text{Tr }\left(\hat{\rho}\,\left(\hat{U}_{0}^{\dagger}\left(T\right)\hat{P}\right)^{n}\hat{Z}_{i}\left(\hat{P}\hat{U}_{0}\left(T\right)\right)^{n}\hat{Z}_{j}\right)\nonumber \\
                        & =\left(-1\right)^{n}C_{ij}^{0}\left(nT\right),
\end{align}
where $\hat{Z}_{i}$ are the $\hat{\sigma}_{j}^{z}$ Pauli matrices and $C_{ij}^{0}\left(nT\right)\equiv\text{Tr }\left(\hat{\rho}\,\left[\hat{U}_{0}^{\dagger}\left(T\right)\right]^{n}\hat{Z}_{i}\hat{U}_{0}^{n}\left(T\right)\hat{Z}_{j}\right)$. The dynamics of $C_{ij}\left(nT\right)$ has subharmonic oscillations characterizing a DTC as long as the dynamics of $C_{ij}^{0}\left(nT\right)$ is a time-liquid with $\lim_{n\to\infty}\lim_{L\to\infty}C_{ij}^{0}\left(nT\right)=c_{ij}\neq0$. The dynamics of $\hat{U}_{0}\left(T\right)$, therefore must be non-ergodic, since for an ergodic dynamics, by definition, $\lim_{n\to\infty}\lim_{L\to\infty}C_{ij}^{0}\left(nT\right)=\left\langle \hat{Z}_{i}\right\rangle \left\langle \hat{Z}_{j}\right\rangle =0$.

We also require the phase to be stable to local periodic rotations of the spins of the form $\hat{R}\left(\left\{ \theta\right\} _{i}\right)\equiv\exp\left[i\sum_{j=1}^{L}\frac{\theta_{i}}{2}\hat{X}_{j}\right]$, which commutes with $\hat{P}$. For simplicity, in this work we will consider only global rotations, namely $\theta_{i}=\theta$. We will denote the perturbed version of $\hat{U}_{0}\left(T\right)$ as $\hat{U}_{\theta}\left(T\right)=\hat{R}\left(\theta\right)\hat{U}_{0}\left(T\right)$.

The stability of the DTC phase therefore builds on the stability of the corresponding \emph{non-ergodic time-liquid} phase supported by $\hat{U}_{\theta}\left(T\right)$.

\begin{figure}[t]
  \includegraphics[width=1\columnwidth]{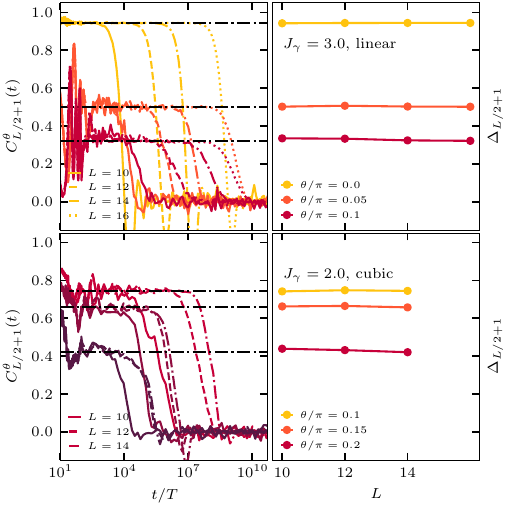}

  \caption{\label{fig:intermediate-plateau}
  \emph{Upper row}: Linear ferromagnetic interaction $J_{\gamma}\left(i-\frac{L}{2}\right)$ with $J_{\gamma}=3$; \emph{Lower row}: Cubic ferromagnetic interaction $J_{\gamma}\left(i-\frac{L}{2}\right)^3$  with $J_{\gamma}=2$.
  \emph{Left column}. Correlation function
  $C_{L/2+1}^{\theta}\left(t\right)$ versus time for a number
  of system sizes (varying line style) and perturbation strengths (varying
  color intensity). The dash--dotted black lines correspond
  to $\Delta_{L/2+1}$, the height of the intermediate plateau. \emph{Right
    column}. The height of the intermediate plateau, $\Delta_{L/2+1}=2^{-L}\sum_{\alpha}\max_{\beta}\left|\left\langle \alpha\left|\hat{Z}_{L/2+1}\right|\beta\right\rangle \right|^{2}$,
  as a function of the system size, for a number of perturbations strengths
  (see legend). The rest of the parameters used in the calculation are
  $h=0.5$, $J_{0}=1$, , $V=0.5$ and $T=1$.}
\end{figure}

It is important to note that, due to Eq.~(\ref{eq:fluctuation-bound}), $\lim_{L\to\infty}\lim_{n\to\infty}C_{ij}\left(nT\right)=C_{ij}^{\theta}\left(nT\right)=0$ in general \footnote{This is true as long as the spectrum of $\hat{U}_{\theta}\left(T\right)$ and $\hat{U}\left(T\right)$ has non-degenerate gaps, which is true in general for a finite system. In addition the spectrum is non-degenerate as we require stability to local periodic perturbations.}. Since time-liquid and DTC phases require this limit to be finite, we conclude that the the limits $t\to\infty$ and $L\to\infty$ should not commute. The most natural way this can happen is that for a \emph{finite} system, $C_{ij}^{\theta}$ develops a non-zero intermediate plateau up to some time, after which it decays to a terminal plateau at 0. The time at which this decay from the intermediate plateau occurs increases with system size. See the left panels of Fig.~\ref{fig:intermediate-plateau} for two examples.

What mechanism can lead to this behaviour? Writing $C_{ij}^{\theta}\left(nT\right)$ in terms of of the eigenvectors of $\hat{U}_{\theta}\left(T\right)$ \footnote{The diagonal terms $\beta=\alpha$ are excluded from the sum because when the eigenstates of $\hat{U}_{\theta}\left(T\right)$ are non-degenerate they are also eigenstates of $\hat{P}$, and $\hat{P}\hat{Z}_{i}\hat{P}=-\hat{Z}_{i},$ such that $\left\langle \alpha\left|\hat{Z}_{i}\right|\alpha\right\rangle =0$.}, $C_{ij}^{\theta}\left(nT\right)=2^{-L}\sum_{\alpha\neq\beta}\left\langle \alpha\left|\hat{Z}_{i}\right|\beta\right\rangle \left\langle \beta\left|\hat{Z}_{j}\right|\alpha\right\rangle e^{i\left(E_{\alpha}-E_{\beta}\right)nT}$   and taking a \emph{finite-time} average up to time $t_{*}$ gives the following estimate for the height of the intermediate plateau,
\begin{equation}
  C_{ij}^{\theta}\left(t\sim t_{*}\right)\sim2^{-L}\sum_{\left|\alpha-\beta\right|\leq\Delta E}\left\langle \alpha\left|\hat{Z}_{i}\right|\beta\right\rangle \left\langle \beta\left|\hat{Z}_{j}\right|\alpha\right\rangle ,\label{eq:inter_plateau_value}
\end{equation}
where $\Delta E\equiv2\pi/t_{*}$ (see~\cite{SuppMatLongTimeLimit} for more details). Since we want the plateau to extend to infinite times in the thermodynamic
limit, we require $t_{*}$ to increase with $L$, and correspondingly
for $\Delta E$ to decrease with $L$. It however cannot decrease
faster than the typical spacing between the quasi-energies, $\delta E=2\pi\cdot2^{-L}$,
since if the sum in Eq.~(\ref{eq:inter_plateau_value}) includes
only the term, $\left\langle \alpha\left|\hat{Z}_{i}\right|\alpha\right\rangle \left\langle \alpha\left|\hat{Z}_{j}\right|\alpha\right\rangle $,
it vanishes. In this work we set $\Delta E$ to be of the order of
a few quasi-energy spacings, $\Delta E=a\delta E$. There will be
an intermediate plateau up to time $t\sim t_{*}=O\left(2^{L}\right)$,
if the sum in Eq.~(\ref{eq:inter_plateau_value}) remains finite
as $L$ increases. For $i=j$, the sum can be bounded by, $\max_{\left|\alpha-\beta\right|\leq\Delta E}\left|\left\langle \alpha\left|\hat{Z}_{i}\right|\beta\right\rangle \right|^{2}.$
Namely, we are looking for systems with off-diagonal matrix elements
which do \emph{not} decay with the system size, and as such violate
the eigenstate thermalization hypothesis (ETH). We designate by $\ket{\alpha}$
and $\ket{\bar{\alpha}}$ the pair of eigenstates for which $\left|\left\langle \alpha\left|\hat{Z}_{i}\right|\beta\right\rangle \right|^{2}$
is maximal and $\left|\alpha-\bar{\alpha}\right|\leq\Delta E\sim2^{-L}$.
Having states $\ket{\alpha}$ and $\ket{\bar{\alpha}}$ localized,
in the sense that they have exponentially decaying two-point correlators
of local observables, is not sufficient, since states for which $\left|\left\langle \alpha\left|\hat{Z}_{i}\right|\bar{\alpha}\right\rangle \right|^{2}$
doesn't decay with system size, will typically be far in energy~\footnote{For example, the eigenstates of the disordered Ising model, $\hat{H}=\sum_{i}J_{i}\hat{Z}_{i}\hat{Z}_{i+1}+h\sum_{i}\hat{Z}_{i}$,
are trivially localized, but eigenstates for which \textbf{$\left|\left\langle \alpha\left|\hat{Z}_{i}\right|\bar{\alpha}\right\rangle \right|$}
is finite are at least $J_{i}+h$ apart in energy.}. Therefore, Anderson insulators or MBL systems do not satisfy this
criteria.

Natural candidates are localised systems where one can find a \emph{local} operator $\hat{M}$ which anti-commutes with $\hat{P}$, but commutes with $\hat{U}_{\theta}\left(T\right)$ in the thermodynamic limit. We will call $\hat{M}$ the Majorana operator. Then $\ket{\bar{\alpha}}=\hat{M}\ket{\alpha}$ will generate a locally similar state, with opposite parity and the same energy up to an exponential correction,such that the entire spectrum is composed of pairs of \emph{quasi-degenerate} states~\citep{Huse:2013bw},  separated by $\sim\exp\left(-L\right)$. For $\left|\left\langle \alpha\left|\hat{Z}_{i}\right|\bar{\alpha}\right\rangle \left\langle \bar{\alpha}\left|\hat{Z}_{j}\right|\alpha\right\rangle \right|$ to remain finite in the thermodynamic limit, the eigenstates $\ket{\alpha}$ should have a system-size independent matrix product (MPS) state representation, which is naturally the case if $\hat{U}_{\theta}\left(T\right)$ is Floquet-MBL. At the same time, the quasi-degeneracy also ensures that the extent of the intermediate plateau as given by Eq.~(\ref{eq:inter_plateau_value}) is up to exponentially long times, $t_{*}\sim\Delta E^{-1}=O\left(2^{L}\right)$.

To summarize, existence of a stable non-ergodic time-liquid and DTC requires a Floquet-MBL system, $\hat{U}_{\theta}\left(T\right)$ with the property $\left[\hat{U}_{\theta}\left(T\right),\hat{P}\right]=0$ and a quasi-degenerate paired spectrum. To be precise, the above conditions are not sufficient, since they do not guarantee that the fluctuations around the \emph{intermediate} plateau of $C_{ii}^{\theta}$ should also vanish in the thermodynamic limit. In what follows we will check for all conditions numerically for a concrete  model.

\emph{Model}.---We use an interacting spin-chain described by the
Hamiltonian
\begin{align}
  \hat{H}_{I} & =\sum_{j=1}^{L-1}\left(J_{j}\hat{Z}_{j}\hat{Z}_{j+1}+V\hat{X}_{j}\hat{X}_{j+1}\right)+h\sum_{j=1}^{L}\hat{X}_{j}\label{eq:tfim-1}
\end{align}
where $\hat{X}_{j}$ and $\hat{Z}_{j}$ are the Pauli matrices $\hat{\sigma}_{j}^{x}$ and $\hat{\sigma}_{j}^{z}$, respectively, $h$ is a transverse field and $J_{i}$ and $V$ are spin-spin couplings. The system is perturbed periodically so that the one-period propagator is $\hat{U}_{\theta}\left(T\right)=\hat{R}\left(\theta\right)\exp\left(-i\hat{H}_{I}T\right)$. We will take $T=1$ throughout and an unbounded ferromagnetic coupling,
\begin{equation}
  J_{i}=J_{0}+J_{\gamma}\left(i-\frac{L}{2}\right)^\alpha,
  \label{eq:Ji}
\end{equation}
with $J_{0}=1$ and $\alpha\geq 1$.

For $V=h=\theta=0$ the Hamiltonian $\hat{H}_{I}$ is diagonal in the basis of eigenstates of $\hat{Z}_{j}$ ($\ket n$, the computational basis), and it is easy to check that for an infinite temperature initial state, $C_{ij}^{\theta}\left(nT\right)=\delta_{ij}$, such that it is a trivial non-ergodic time-liquid. Flipping the spins periodically (that is, $\theta=0$) yields the corresponding DTC phase. Since for the infinite temperature initial state the correlation function is zero for $j\neq i$ so that there is no spatial order, in what follows we only consider the case $j=i$ when studying the regime away from $V=h=\theta=0$.

\emph{Quasi-degeneracy.}---At the special point $V=h=\theta=0$ the spectrum of $\hat{H}_{I}$ is degenerate: $\ket n$ and $\hat{P}\ket n$ have the same eigenvalue. Any $\hat{Z}_{i}$ anti-commutes with $\hat{P}$ and commutes with $\hat{H}_{I}$ so that the states $\ket{\pm}=\left(\ket n\pm\hat{P}\ket n\right)/\sqrt{2}$ have the property $\ket{\pm}=\hat{Z_{i}}\ket{\mp}$ required from the Majorana operator, $\hat{M}$. Since $\ket n$ and $\hat{P}\ket n$ are \emph{globally} different, any \emph{local} perturbation $\lambda\hat{H}_{\text{V }}$(we take $\lambda>0$) will couple them only at order $O\left(L\right)$ of the perturbation theory, such that $\lambda^{L}\left\langle n\left|\hat{H}_{\text{loc}}^{L}\hat{P}\right|n\right\rangle \neq0$. This will result in an exponential splitting between the $\ket{\pm}$ states $\left(\lambda/J\right)^{L}$, if $\left|J\right|>\lambda$, where for simplicity we assumed $J_{i}=J$.

To generalise the argument to a non-uniform $J_{i}$, consider a partitioning of the system into regions where $\left|J\right|/\lambda<1$ and $\left|J\right|/\lambda>1$. Labelling the regions by integers $r$, successive $r$ have $\left|J_{i}\right|/\lambda$ larger/smaller than 1. Since interactions are local, the total energy will be the sum of the energies in each region (up to exponentially small corrections that we neglect).  The total energy then is $E\left[\left\{ n_{r}\right\} \right]=\sum_{r}\epsilon_{n_{r}}^{(r)}$ where $n_{r}$ labels the state inside each region. For regions $r$ where $\left|J_{i}\right|<\lambda$, the $\epsilon_{n_{r}}^{(r)}$ the quasi-degeneracy is completely lifted. However, for regions where $\left|J_{i}\right|>\lambda,$ the lifting is only by $\sim\exp\left(-L_{r}\right)$ where $L_{r}$ is the spatial extend of the region. Given an eigenstate with energy $E\left[\left\{ n_{r}\right\} \right]$, we can construct another eigenstate with energy  $E\left[\left\{ n_{r}'\right\} \right]$, which is separated from it by $\sim\exp\left(-L_{r}\right)$, by replacing a state $n_{r}$ in region $r$ by its quasi-degenerate partner. The smallest possible separation will be achieved by taking the largest regions with where $\left|J_{i}\right|>\lambda$.  Thus the system is quasi-degenerate, as long as there is a region with $\left|J_{i}\right|>\lambda$ that grows with system size, which  is certainly the case for Eq.~\ref{eq:Ji} with $\alpha\geq 1$. For more details on the quasi-degeneracy see~\cite{SuppMatQuasiDegeneracy}.

\emph{Localization}.---The model with $V=h=\theta=0$ has trivially localized domain-walls (DWs). In what follows we show that localization remains robust, as long as the interaction $J_{i}$ is unbounded.  We will only consider the stability for $\theta=0$, the stability for $\theta\neq0$ follows from general arguments of stability of Floquet-MBL. We write the total Hamiltonian $\hat{H}_{I}$ as $\hat{H}_{I}=\hat{H}_{0}+\hat{H}_{h}+\hat{H}_{V}$, where $\hat{H}_{h,V}$ are the two off-diagonal terms in Eq.~(\ref{eq:tfim-1}).  An eigenstate of $\hat{H}_{0}$ with $n$ DWs at positions $i_{1},\ldots i_{n}$ has energy $E_{\left\{ i_{m}\right\} }=\sum_{j}J_{j}-2\sum_{i\in\left\{     i_{m}\right\} }J_{i}$.  Both of $\hat{H}_{h}$ and $\hat{H}_{V}$ only connect it to states with $n,n\pm2$ DWs~\footnote{Except when   acting at the edges of the system when they may also connect to   states with $n\pm1$.}. Adding or removing a DW at $i$ changes $E_{\left\{ i_{m}\right\} }$ by $\pm2J_{i}$, while moving a DW from $i$ to $i+r$, changes the energy by $2\left(J_{i}-J_{i+r}\right)$. Restricting our discussion to $\hat{H}_{h}$ for simplicity (the discussion for $\hat{H}_{V}$ is almost identical), we note that given an eigenstate of $\hat{H}_{0}$ with $n$ DWs, the states closest in energy connected to it by leading-order perturbation theory also have $n$ DWs, with a single DW moved by one site. Restricting the Hamiltonian to single domain-wall states, valid for small $h$ (see~\cite{SuppMatDWs} for more details), and writing $\ket i$ for a state with a single domain wall at $i$, gives the effective single-particle Hamiltonian \footnote{For $h=V=0$, there   are two degenerate states with the domain wall at $i$. One can label   them with an internal pseudospin degree of freedom,   $\ket{i,\uparrow/\downarrow}$ and the Hamiltonian in   Eq.~\ref{eq:hopping-fm} would then not mix the two. We omit this   here for notational simplicity.}
\begin{equation}
  \bra i\hat{H}\ket j=2J_{j}\delta_{i,j}+h\left(\delta_{i,j+1}+\delta_{i,j-1}\right),\label{eq:hopping-fm}
\end{equation}
so that a DW behaves like a single particle hopping around with a uniform hopping amplitude $h$ and potential $J_{j}$. For $\alpha=1$ the problem reduces to Stark localization~\citep{zener_theory_1934,Wannier1960} (see Eq.~\ref{eq:Ji}). Localization persists for states with multiple DWs (see~\cite{SuppMatDWs}) or power $\alpha>1$.

\emph{Correlation function.---}The combination of localization and quasi-degeneracy leads to each eigenstate $\ket{\alpha}$ having a partner eigenstate $\ket{\bar{\alpha}}$ lying within a quasi-energy window $\Delta E=O\left(e^{-L}\right)$ and having an off-diagonal matrix element $\left|\left\langle \alpha\left|\hat{Z}_{i}\right|\bar{\alpha}\right\rangle \right|^{2}$ remaining finite even in the TDL. In this case, the correlation function, $C_{i}^{\theta}$, will have an intermediate plateau of height $\Delta_{i}=2^{-L}\sum_{\alpha}\left|\left\langle \alpha\left|\hat{Z}_{i}\right|\bar{\alpha}\right\rangle \right|^{2}$ for times up to $t_{*}\sim\exp\left(L\right)$ as discussed earlier; after this time, the correlation function decays. This is shown in the left panels of Fig.~\ref{fig:intermediate-plateau} where we numerically compute $C_{L/2+1}^{\theta}$ for two choices of $J_i$ and a number of $\theta$ and system sizes. The left panels show that $C_{L/2+1}^{\theta}$ displays an intermediate plateau persisting for times exponentially large in system size before decaying and fluctuating around zero. In the right panels we display $\Delta_{L/2+1}$ as follows: for each eigenstate $\ket{\alpha}$ of $\hat{U}_{\theta}\left(T\right)$ we take the 10 closest (in quasi-energy) $\ket{\beta}$, and compute $\Delta_{L/2+1}=2^{-L}\sum_{\alpha}\max_{\beta}\left|\left\langle \alpha\left|\hat{Z}_{L/2+1}\right|\beta\right\rangle \right|^{2}$. The dash-dotted lines in the left panels show that the $\Delta_{L/2+1}$ obtained quantitatively reproduces the height of the intermediate plateau, while the right panels show that this height doesn't decay with system size for a range of perturbations $\theta$.

We now turn to the fluctuations of $C_{i}^{\theta}\left(t\right)$ around the intermediate plateau, showing that they vanish in the thermodynamic limit: $\lim_{t\to\infty}\lim_{L\to\infty}C_{i}^{\theta}\left(t\right)=c>0$.

\begin{figure}[t]
  \includegraphics[width=1\columnwidth]{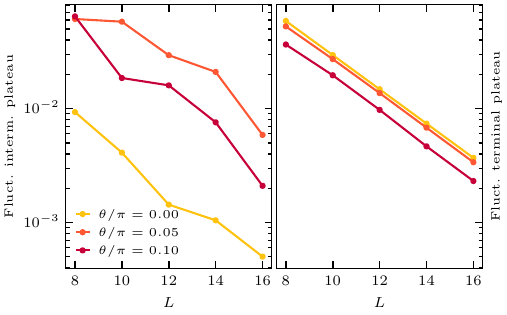}

  \caption{\label{fig:fluctuations-noflip}Temporal fluctuations of the correlation
    function $C_{L/2+1}^{\theta}\left(t\right)$ around the intermediate
    plateau (left) and the terminal plateau (right) as a function of the
    system size and perturbation strength $\theta$ for a linear potential,
    $\alpha=1$. The rest of the parameters
    used in the calculation are $h=0.5$, $J_{0}=1$, $J_{\gamma}=3$,
    $V=0.5$ and $T=1$.}
\end{figure}
\begin{acknowledgments}
  \emph{Fluctuations}.---From here on we focus on the case $\alpha=1$. The magnitude of the temporal fluctuations in the terminal plateau can be calculated by infinite-time average of $\left|C_{\theta}\left(t\right)\right|^{2}$ and is given by $2^{-L}\sum_{\alpha,\beta}\left|\left\langle \alpha\left|\hat{Z}_{i}\right|\beta\right\rangle \right|^{4}$. As discussed bellow Eq.~(\ref{eq:fluctuation-bound}) it decays exponentially with $L$, as we indeed see in the right panel of Fig.~\ref{fig:fluctuations-noflip}. To compute the temporal fluctuations around the intermediate plateau, we compute the time-average up-to time $t_{*}$ which we explicitly set as the departure time from the intermediate plateau. For all studied values of parameters the fluctuations decay with system size. Combined with the fact that $t_{*}\sim\exp\left(L\right)$, this suggests that $\lim_{t\to\infty}\lim_{L\to\infty}C_{i}^{\theta}\left(t\right)=c>0$, namely there exists a stable non-ergodic liquid phase. Adding a spin-flipping operator $\hat{P}$ every period, the non-ergodic time-liquid is readily modified into a DTC, and shows subharmonic oscillations for sufficiently large $J$ (see Fig.~\ref{fig:short-time-dynamics}).

  \emph{Conclusions.}--In conclusion, we have presented a detailed study of DTCs in a class of non-disordered, non-integrable Ising-type systems. We discuss a general mechanism, which relies on localisation and quasi-degeneracy. We then present approximate analytical arguments that our models do have these properties, with localisation relying the physics of the domain walls and perturbatively stable degeneracy lifted by finite-size effects resulting in finite lifetime of the DTC for finite-sized systems. We finally numerically demonstrate these results for two concrete examples, one of which is inspired by the usual Stark problem.

  We note that the mechanism we propose is not limited to the specific models we study, but is a feature a general class of Floquet-MBL systems; the particular example of a linear potential has also been studied in Refs.~\cite{Kshetrimayum2020b,Liu2023}.

  Overall we note that DTC order is stable the thermodynamic limit, while the limits of long time and large size do not commute--a telltale sign that the mechanism we discuss is at play, as the finite-size lifting of the degeneracy is the root cause of the finite lifetimes.

  Our class of models is disorder-free, unitary, many-body, and display DTC with exponentially-long lifetimes. Experimental demonstration of our results requires similar machinery as that of disordered DTCs~\citep{Khemani2016,Mi2022a} and so should be within reach of current setups.
\end{acknowledgments}

\bibliography{ms,yb, sm}
\end{document}